\documentclass{article}
\usepackage{authblk}

\usepackage{booktabs} % For formal tables
\usepackage{ifpdf}

\usepackage{cite}

\usepackage{amsmath}

\usepackage{algorithmic}

\usepackage{array}

\usepackage{url}

  \usepackage[pdftex]{graphicx}

\begin{document}

\title{REOH: Using Probabilistic Network for Runtime Energy Optimization of Heterogeneous Systems (IFI-UiT Technical Report 2018-81)}

\author{Vi Ngoc-Nha Tran}
\author{Tommy Oines}
\author{Alexander Horsch}
\author{Phuong Hoai Ha}
\affil{Department of Computer Science\\
UiT The Arctic University of Norway\\
Tromso, Norway\\
\{vi.tran, tommy.s.oines, alexander.horsch, phuong.hoai.ha\}@uit.no}

\date{September 16, 2018}
\maketitle

\thispagestyle{empty}

\begin{abstract}
Significant efforts have been devoted to choosing the best configuration of a computing system to run an application energy efficiently. However, available tuning approaches mainly focus on homogeneous systems and are inextensible for heterogeneous systems which include several components (e.g., CPUs, GPUs) with different architectures.

This study proposes a holistic tuning approach called REOH using probabilistic network to predict the most energy-efficient configuration (i.e., which platform and its setting) of a {\em heterogeneous} system for running a given application. Based on the computation and communication patterns from Berkeley dwarfs, we conduct experiments to devise the training set including 7074 data samples covering varying application patterns and characteristics. Validating the REOH approach on heterogeneous systems including CPUs and GPUs shows that the energy consumption by the REOH approach is close to the optimal energy consumption by the Brute Force approach while saving 17\% of sampling runs compared to the previous (homogeneous) approach using probabilistic network. Based on the REOH approach, we develop an open-source energy-optimizing runtime framework for selecting an energy efficient configuration of a heterogeneous system for a given application at runtime.
\end{abstract}

\section{Introduction}
\label{introduction}
%Motivation of energy-efficiency for HSA
%Motivation of adaptive tuning approaches. 
%Why is it hard? (E.g., why do naive approaches fail?)
%How this study differs?
%Summary of contributions
%Motivation of energy-efficiency for HSA
Improving the energy efficiency and reducing energy consumption are ones of the most important requirements of computing systems. %Nowadays, Heterogeneous System Architecture (HSA) is widely introduced to achieve better efficiency regarding power and performance 
%programmability and portability 
%of applications. HSA combines a wide range of devices, varying from CPU to GPU and mobile devices. 
%Motivation of adaptive tuning approaches. 
The factors that have impacts on the application performance and its optimization strategies are algorithm design and implementation (i.e., control flow, memory types, memory access pattern and instruction count) and its execution configuration \cite{Feng:2012:ODW:2188286.2188341}. When an application runs on a heterogeneous system, one of the strategies to reduce energy consumption is to run the application with an appropriate system configuration.

%Based on %https://arxiv.org/ftp/arxiv/papers/1412/1412.1297.pdf, the impact factors are total execution time, class of application, problem size, programming language, I/O time, memory access time, CPU access time
%Applications can change class when programming language changes from OpenMP to OpenCL. 
%Moreover, when problem size increases, an application can change from computation bound to memory bound
%-> motivation to tune/choose platform/configuration based on input size and sample execution of application (implementation/programming language)
%3. Why is it hard? (E.g., why do naive approaches fail?)
Several attempts \cite{
%ATLAS2000, LAPACK, SPIRAL, OSKI2005, 6877283, 
OPENCL, Powercap, POET, Mishra:2015, 6413638, 6877247, Nath:2015:CPM:2830772.2830826, 7920860, 8327055, 6337630, Wang:2017:GPE:3152042.3152066} have been made to find the best configurations to run an application to achieve energy efficiency. However, available tuning approaches are mostly conducted for homogeneous systems while little research considers heterogeneous systems including several platform components (e.g., CPUs and GPUs) with different types of processing units and different architectures.

%Energy-optimizing approaches for homogeneous systems are only applicable for an {\em individual} component (e.g., CPUs or GPUs) of a heterogeneous system and therefore they cannot be used as a {\em holistic} approach for a heterogeneous system consisting of different components (e.g., CPUs and GPUs). Holistic approaches that consider all heterogeneous components together, are essential for optimizing the energy consumption of the whole heterogeneous system.
%(More explanation about how the latest approaches works for homogeneous systems.)
%The available approaches to improving the energy-efficiency of heterogeneous systems are mostly devoted to the hardware design. The methodology to tune configurations for energy optimization on heterogeneous systems, however, is missing. 
Table \ref{table:auto-tuning-framework} summarizes the related work to this study according to the four aspects: the optimization goal (i.e, Optimization), whether the optimization object is configuration or code variant (i.e., Object), whether the targeted system is homogeneous or heterogeneous (i.e., System), and whether the approach is applicable for general or specific applications (i.e., Application). 
%In the context of this study, heterogeneous systems are the system containing different types of processing units (e.g., CPU and GPU are different types of processing units). 
Table \ref{table:auto-tuning-framework} shows how our study is different from its related work. The goal is to optimize energy efficiency by choosing an appropriate configuration of heterogeneous systems for a given application. The details of the related work are described in Section \ref{related_work}.
\begin{table}
	%\resizebox{0.5\textwidth}{!}{\begin{minipage}{\textwidth}
	\caption{Auto-Tuning Framework}
	\label{table:auto-tuning-framework}
	\centering
	\small
	%\resizebox{\columnwidth}{!}{%
	\begin{tabular}{llllllll}
		\hline\noalign{\smallskip}
		\textbf{Study} 	 &\textbf{Optimization} 	&\textbf{Object}  	&\textbf{System}  	 &\textbf{Application} \\						
		\noalign{\smallskip}\hline\noalign{\smallskip}						
		%ATLAS, BLAS \cite{ATLAS2000} 	&Time 	&Code variant 	 	     &Homogeneous  	&Specific\\
		%LAPACK \cite{LAPACK} 	& 	&	 	     &(i.e., CPU or GPU)  	&\\
		%SPIRAL\cite{SPIRAL} 	 	& 	&	 	     & 	&\\
		OSKI \cite{OSKI2005} 	&Time 	&Code variant 	     &Homogeneous 	&Specific \\
		& 	&	 	     &(i.e., CPU) 	&(i.e., Sparse kernels)\\
		\noalign{\smallskip}\hline\noalign{\smallskip}				
		Nitro \cite{6877283} 	&Time 	&Code variant 	     &Homogeneous	&General\\
		& 	& 	     &(i.e., GPU)	&\\
		\noalign{\smallskip}\hline\noalign{\smallskip}	
		%OPENCL \cite{OPENCL} 	&Time 	&Compiler-  	     &Homogeneous 	&General\\
		%& 	&factor	 	     &(i.e., GPU) 	&\\
		PowerCap  	&Timeliness 	&Configuration 	     &Homogeneous	&General\\
		\cite{Powercap}&Energy- 	& 	     &(i.e., CPU)	&\\
				&efficiency 	& 	     &	&\\
		\noalign{\smallskip}\hline\noalign{\smallskip}
		POET \cite{POET} 	&Energy- 	&Configuration  	     &Homogeneous	&General\\
		&efficiency 	& 	     &(i.e., CPU)	&\\
		\noalign{\smallskip}\hline\noalign{\smallskip}
		LEO \cite{Mishra:2015} 	&Time &Configuration	     &Homogeneous	&General\\
		&Energy- 	& 	     &(i.e., CPU)	&\\
		&efficiency 	& 	     &	&\\
		\noalign{\smallskip}\hline\noalign{\smallskip}
		%new
		HPC runtime&Energy- 	&Configuration 	     &Homogeneous	&General\\
		framework \cite{6413638}&efficiency 	& 	     &(i.e., CPU)&\\
		\noalign{\smallskip}\hline\noalign{\smallskip}
		GPU models \cite{6877247}&Power 	&Configuration 	     &Homogeneous	&General\\
		& 	& 	     &(i.e., GPU)&\\
		\noalign{\smallskip}\hline\noalign{\smallskip}
		CRISP \cite{Nath:2015:CPM:2830772.2830826}&Energy 	&Configuration 	     &Homogeneous	&General\\
		& 	& 	     &(i.e., GPGPU)&\\
		 \noalign{\smallskip}\hline\noalign{\smallskip}
		MPC \cite{7920860}&Energy- 	&Configuration 	     &Homogeneous	&General\\
		&efficiency 	& 	     &(e.g., GPGPU)&\\
		 \noalign{\smallskip}\hline\noalign{\smallskip}
		GreenGPU \cite{6337630, MA201621}&Energy- 	&Workload division      &Heterogeneous	&Specific \\
		&efficiency 	&Frequency &(e.g., CPU and GPU)&(i.e., Iterative\\
		& 	&  &&applications)\\
		 \noalign{\smallskip}\hline\noalign{\smallskip}
		GPGPU DVFS  \cite{8327055}&Energy- 	&Configuration 	     &Homogeneous	&General\\
		&efficiency 	& 	     &(i.e., GPGPU) &\\
		\noalign{\smallskip}\hline\noalign{\smallskip}
		GPGPU SVR  \cite{Wang:2017:GPE:3152042.3152066}&Energy- 	&Configuration 	     &Homogeneous	&General\\
		&efficiency 	& 	     &(i.e., GPGPU) &\\
		\noalign{\smallskip}\hline\noalign{\smallskip}
		Market mechanism  	&Service quality& High-level	     &Heterogeneous	&General\\
		\cite{Guevara:2014:MMM:2584468.2541258}&Energy- 	& configurations	     &(e.g., CPUs & \\
				&efficiency 	&(i.e., Datacenters) 	     &and microprocessors)	&\\
		\noalign{\smallskip}\hline\noalign{\smallskip}
		REOH (this study) 	&Energy- 	&Configuration	     &Heterogeneous	&General\\
		& efficiency 	&	     &(e.g., CPU and GPU) 	&\\
		\noalign{\smallskip}\hline\noalign{\smallskip}
	\end{tabular}
\end{table}

%4. Why hasn't it been solved before? (Or, what's wrong with previous proposed solutions? How does mine differ?)
The main goal of existing tuning approaches 
%\cite{ATLAS2000, LAPACK, SPIRAL, OSKI2005, 6877283} 
is to improve energy-efficiency.
% \cite{POET, Powercap, Mishra:2015}. 
%Energy consumption requires fine-grained sensors for measurement and new programming libraries that are only available in the modern high performance devices. Measuring energy of computing systems are complicated to be done manually and requires automatic tools as well as good prediction models.
However, the existing models are mostly built for homogeneous systems, which has only one type of devices such as GPU \cite{6413638, 6877247, Nath:2015:CPM:2830772.2830826, 8327055, 7920860, Wang:2017:GPE:3152042.3152066
%6877283
} or CPU \cite{POET, Powercap, Mishra:2015}. There are also a set of studies \cite{7360199, 7501903, 7863726} for heterogeneous systems (i.e., APUs) but they are mainly focus on improving performance instead of energy-efficiency. 

The existing heterogeneous approaches in the Table \ref{table:auto-tuning-framework} are either for specific applications (i.e., iterative applications that can be divided to several iterations where execution time of the next iteration can be predicted based on the current iteration) \cite{6337630, MA201621} or for finding a heterogeneous balance of datacenter \cite{Guevara:2014:MMM:2584468.2541258} where the configuration at datacenter level is a mix of CPUs or microprocessors.

Among the available tuning approaches, probabilistic model-based approaches have their advantages of not requiring prior knowledge on the targeted application or the throughout understanding of system components like other approaches \cite{Nath:2015:CPM:2830772.2830826, 8327055}. By finding the similarity between the targeted application from sampling data and previous observed applications from training data, it can quickly provide the accurate estimation of energy consumption for the targeted application. 

The previous probabilistic model based approaches only applicable for homogeneous systems (i.e., CPUs). Heterogeneous systems have complex structures containing different platform architectures (e.g., CPUs, GPUs, FPGAs, ASICs) where each platform has its own sets of settings and methods to change its configurations. Applying the probabilistic model based approach \cite{Mishra:2015} on each individual platform of a heterogeneous system requires the analysis of the available settings and a new configuration data for each platform. In the other words, it requires separated sets of training and sampling data, and separated runs of prediction for each platform. This results in more sampling runs than doing one prediction for a heterogeneous system with only one whole set of training and sampling data. Therefore, the probabilistic model based approaches for heterogeneous systems requires the analysis of the available settings of all included platforms within a heterogeneous system and finding the setting equivalence of one platform to another platform.

In this study, we propose a way to unify the configurations of different platforms on a heterogeneous system and do the prediction only once. This way we save energy of the sampling runs. Even though we evaluate the probabilistic model-based approach (i.e., REOH) on a system containing CPU and GPU only, REOH is general for heterogeneous systems which contain any architectures (e.g., CPUs, GPUs, FPGAs, ASICS) where we can identify and change their configurations (i.e., the combination of number of cores, memory and frequency) in runtime.

The proposed approach aim to address the following research question: {\em"Given executable files of an application and a heterogeneous system containing platforms with different architecture, which system configuration (i.e., platform and its setting) to run the application most energy-efficiently?"}

This study propose holistic tuning approach based on probabilistic model to predict the most energy-efficient configuration of heterogeneous systems for a given application. Based on the application communication and computation patterns (i.e., Berkeley dwarfs \cite{Asa06}, we choose the Rodinia benchmarks \cite{Rodinia} for the experiments and devise a training data set. The objectives when choosing the benchmarks are to devise a training data set that cover a wide range of application patterns and characteristics. %The validation of the approach shows that the proposed approach selects the more energy-efficient configuration compared to the homogeneous approach \cite{Mishra:2015} and the energy consumption is closer to the optimal energy consumption by the brute-force approach. 

We also provide an open-source energy-optimizing runtime framework to choose which configuration of a heterogeneous system to run a given application at runtime. Even though the open-source is for the experimented system including only one CPU and one GPU, the code is available and can be adjusted to heterogeneous systems containing other types of platforms as long as changing platform configurations during runtime is supported.

% Study limitation
This study is for applications that runs on one platform (e.g., CPU or GPU) at a time. The application has different executable files for different platforms (e.g., CPU or GPU) that can be chosen during runtime. For example, Rodinia benchmarks suite \cite{Rodinia} supports programming models such as OpenCL which can provide different executable files of the same benchmark. This approach, however, 
%does not consider the applications that run on multiple platforms in the same execution. 
can also apply to applications that can be divided to several phases. Each phase is wrapped in an executable file and can be considered as one application in REOH approach. Therefore, each phase of such applications only runs on one platform but the whole execution with different phases runs on several platforms.

%This approach has overhead as the prediction process and the application executions on sample configurations. In order to enable energy saving, this approach targets applications with long running time or having many repeated instances as well as having phases to change configurations at runtime.
%more explaination here
%-built for heterogeneous systems by considering both static and dynamic power of each platform included.

In this work, the following contributions have been made.
\begin{itemize}
	\item Devise a new holistic tuning approach for heterogeneous systems using probabilistic network, which is called REOH. In this study, we propose a method to unify the configurations of different platform types (e.g., CPU and GPU), consider the total energy of both static and dynamic energy and devise a training data set containing 7074 samples by running a selected set of 18 applications based on the knowledge of application patterns from Berkeley dwarfs on a total of 393 system configurations. %To the best of our knowledge, REOH is the first approach that can tune the configuration of a heterogeneous system for general applications to optimize their energy consumption (cf. Table \ref{table:auto-tuning-framework}).    
	\item Validate the REOH approach on a heterogeneous system consisting of CPU and GPU, showing that %REOH approach achieves better results than the previous probabilistic model approach for homogeneous systems \cite{Mishra:2015}) on each individual platform. In most of the cases (e.g., 17 out of 18 applications), 
	REOH approach achieves the close energy consumption (i.e., within 5\% different) to the optimal energy consumption by the brute-force approach when choosing the most energy-efficient system configuration for the applications while saving 17\% number of sampling runs than the existing probabilistic network approaches \cite{Mishra:2015}.
	%we measure the energy consumption on 2 types of HPC system, such as CPU and GPU for chosen Rodinia benchmarks. The developed framework improve energy saving compared to existing approaches (LEO) considering only cpu data.
	\item Develop an open-source energy-optimizing runtime framework for selecting an energy efficient configuration of a heterogeneous system for a given application at runtime. The framework takes as the input the executable files that the users want to run on a targeted heterogeneous system. Then the framework will choose an appropriate configuration of the targeted heterogeneous system to run the executable files energy-efficiently. This tool is %validated and 
	provided as an open source for scientific research purposes. 
\end{itemize}

The content of the paper is organized as follows. Section \ref{tuning_approach} describes REOH, the energy optimization approach for heterogeneous systems. %Section \ref{training_data} explains how to devise the training data set. 
In Section \ref{energy_saving}, we validate the approach on a heterogeneous system consisting og CPUs and GPUs. Based on the proposed energy optimization approach, Section \ref{runtime_app} describe the energy-optimizing runtime framework and its implementation. The related work to this study is summarized in Section \ref{related_work}. Section \ref{conclusion} concludes the study. 

\section{A Holistic Tuning Approach for Heterogeneous Systems}
\label{tuning_approach}
This section first explains the theory behinds the probabilistic graphical model-based approach \cite{Mishra:2015}, and then describe the improvements of REOH, the holistic tuning approach enhanced for heterogeneous systems. 
\subsection{Probabilistic Network Approach}
\label{Probabilistic Network Approach}
%Do we need a seperate section to explain this approach?
The REOH approach is based on the probabilistic graphical model-based approach \cite{Mishra:2015} to predict the energy consumption of all configurations in the systems from the offline training data and the online sample data. The probabilistic graphical model-based approach use hierarchy directed Bayesian network to exploit the conditional dependence of unobserved variables to the previous observed applications. In the context of this study, the unobserved variables are the power consumption and execution time of a new application on every system configurations that we want to estimate. The previous observed applications is the training data. Therefore, the modeling approach using probabilistic network requires no prior knowledge on the targeted application or low-level modeling details of system components (e.g., modeling power of each platform instruction unit or component).
 % Probabilistic model
%????Explain the model behind?????

 Offline training data is the execution times and power consumption of a set of applications. Power is computed by dividing the measured energy consumption by the measured execution time. The offline training data is collected by executing a set of applications that capture a wide range of computation and communication patterns on all possible configurations of the system. The online sample data is the execution time and power consumption of a targeted application when running on some selected configurations of the system. The offline training data provides the knowledge on representative applications while the online sample data provides the knowledge of the targeted application. Base on both offline training and online sample data, a probabilistic graphical model predicts the data for the targeted application on all available configurations of the systems. Based on the prediction results, the best configurations for energy consumption of the targeted application is obtained.
 
\begin{figure}[htp]
	\centering
	\includegraphics[width=0.5\textwidth]{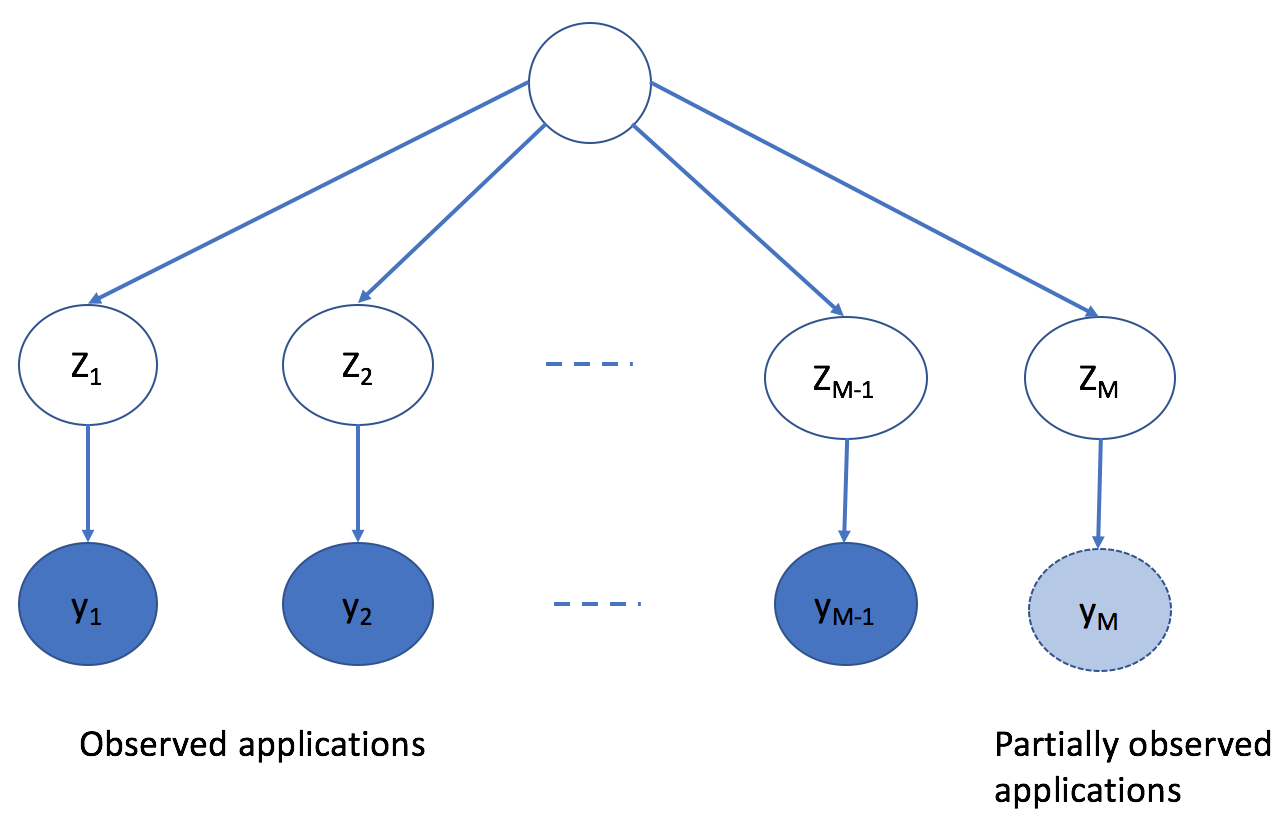}
	\caption{Bayesian Model}
	\label{Fig:Bayesian}
\end{figure}

 The graphical models in this approach uses Bayesian networks (e.g., directed graphical models), to capture the dependence between random variables. In this network, a node is a random variable and the directed edge connecting two nodes is the dependence of the variables. The nodes are either hidden (unknown) or observed (known). %The Bayesian networks with several layers of hidden nodes are hierarchical networks. 
 The Bayesian model in this approach is drawn in Figure \ref{Fig:Bayesian}, where $y_{M}$ is the partially unobserved application (i.e., the applications whose only measurement of sample run are known) that need to be estimated; $y_{1}$ to $y_{M-1}$ is fully observed applications (i.e., training data); and blank nodes $Z_{i}$ and their root are the hidden nodes. 

First the model estimates the missing values by a regression technique. Then, the Expectation Maximization algorithm in statistic is applied to estimate both power and performance of each partially unobserved application on all configurations. The algorithm is the iteration of expectation (E) and maximization (M) function until convergence. The E step is a function computing the expectation of the log-likelihood from the current estimated parameters. From the log-likelihood found in the E step, The M function computes parameters maximizing the expected likelihood. The probabilistic graphical model-based approach \cite{Mishra:2015} iterates E and M step until the convergence to obtain the estimated parameters. The more details of the proof can be found from \cite{Mishra:2015}. 

 %targeted which application 
The probabilistic graphical models \cite{Mishra:2015} targets applications which have long running time or many repeated instances as well as applications have phases to change configuration online. Energy consumption of such applications can be reduced by using probabilistic graphical model-based approach.
% This approach has overhead as the prediction process and the application execution on sample configurations. This approach targets applications with long running time and operate at a range of different utilization to enable energy saving.
\subsection{Unifying platform configurations }
Unlike the previous (homogeneous) probabilistic graphical models approach \cite{Mishra:2015}, the REOH approach proposed in this study is for heterogeneous systems including different platforms with different architectures. The probabilistic modeling approach requires experimental data from a set of configurations that can be tuned during runtime. %Some hardware do not have the sufficient number of tuning configurations and therefore, the probabilistic graphical model approach \cite{Mishra:2015} is not applicable for every individual platform. For instance, the GPU used for the experiments in this study (e.g., Nvidia Quadro K620) have less than five configurations. Even though there are only 5 configurations of the experimented GPU, the gap between the optimum and the worst energy consumption of the GPU is huge (i.e., 10 times for application 17) as shown in Figure \ref{fig:GPU_comparison_unsorted_energy}. Moreover, the most energy-efficient configuration for GPU among all applications are varying. Both emphasize the need of finding the best configurations, even for GPU alone. However, with 5 configurations, the experimented GPU does not provide sufficient data to be applied with the probabilistic approach for homogeneous systems. The details of minimum required data size are explained in Section \ref{sec:Approach validation}. Therefore, for heterogeneous systems containing several platforms, the uniform of the platform configurations is required. 
% \begin{figure}[htp]
%	\centering
%	\includegraphics[width=0.5\textwidth]{diagrams/GPU_comparison_unsorted_energy.jpg}
%	\caption{The optimal and worst energy consumption of GPU}
%	\label{fig:GPU_comparison_unsorted_energy}
%\end{figure}

The configurations must be pre-defined and provided in training data. For REOH, the configurations are the combination of the number of cores, the core frequency and the number of memory controllers. An example of CPU configuration is 24 cores running at frequency 1.7 GHz with two memory channels. Each platform architecture has its own hardware specification with different numbers of cores, the core frequencies or memory controllers \cite{Mishra:2015}. For heterogeneous systems including several platforms with different architectures, in order to apply the probabilistic approach, finding the equivalence of configurations from different platforms is essential.

%For heterogeneous systems containing more advance GPUs, the number of GPU configurations can be numerous when there are supported methods to change their settings (e.g., core frequencies, memory frequencies). Unifying the configurations of different platforms in heterogeneous systems is therefore, essential.
In this section, we propose a methodology to convert the configurations of different platforms.
 %For example, to devise the probabilistic graphical model for heterogeneous systems including both CPUs and GPUs, the configurations of all included platforms need to be unified. The considered hardware features defining the configurations are three numbers: the number of used cores, the number of DVFS settings, and the number of memory controllers \cite{Mishra:2015}. 
%For instance, the considered system has the CPU (2 x Xeon E5-2650Lv3) with 24 cores, 8 DVFS settings and 2 memory controllers, which results as $24 \times 8 \times 2 = 384$ configurations. 
We consider the peak compute flops and peak memory bandwidth when finding the equivalence of the configurations of different platforms. The study by Lee et.al. \cite{Lee:2010:DGV:1815961.1816021} provided a comparison of CPU and GPU performance on 14 kernels considering architectural differences such as processing element (or PE) and bandwidth differences. The average performance (in flops) of each processing element is computed by dividing the platform computing flops by the total number of processing elements in the platform: $Flops_{PE}= \frac{PeakFlops}{TotalPE}$. In the context of this study, the total processing elements are the number of cores available in the platform. E.g., $Flops_{CPUcore}= \frac{PeakFlops_{CPU}}{TotalCores_{CPU}}$ and $Flops_{GPUcore}= \frac{PeakFlops_{GPU}}{TotalCores_{GPU}}$.

Therefore, to unify the number of cores in GPU (or $nGPUcore$) with a equivalent number of cores in CPU (or $nCPUcore$), we compare performance of CPU cores and GPU cores as in Equation \ref{eq:core-convert2}: 
%\begin{equation} \label{eq:core-convert}
%\frac{Flops_{CPUcore}}{Flops_{GPUcore}} = \frac{PeakFlops_{CPU}}{TotalCores_{CPU}} \times %\frac{PeakFlops_{GPU}}{TotalCores_{GPU}}
%\end{equation}
\begin{equation} 
\begin{split}
\label{eq:core-convert2}
	nGPUcore =  \frac{Flops_{GPUcore}} {Flops_{CPUcore}} \times nCPUcore \\
	= \frac{PeakFlops_{GPU}}{TotalCores_{GPU}} \times \frac{TotalCores_{CPU}}{PeakFlops_{CPU}}\times nCPUcore
\end{split}
\end{equation}
In our heterogeneous system, there are two platforms: CPU Xeon E5-2650Lv3 has 24 cores and peak performance as 115.2 GFlops while GPU Nvidia Quadro K620 has 384 cores with peak performance as 860 GFlops. 
%The CPU has 384 configurations while the GPU has 5 configurations. The number of GPU configurations is small because only the number of cores is changeable. 
%removed for CR version 
%The number of cores for GPU in the context of this study is the work group size assigned to the applications. 
The average performance for a CPU core is $\frac{115.2}{24}=4.8$ GFlops while the average performance for a GPU core is $\frac{860}{384}=2.24$ Gflops. One GPU core is equivalent to $\frac{24}{115.2} * \frac{860}{384}=0.47$ CPU core, which is approximately half of the performance of one CPU core. Therefore, one GPU core is approximately equivalent to 0.5 CPU core. 

Similarly, we convert the number of memory controllers of GPU (or $nGPUmem$) to the number of memory controllers in CPU (or $nCPUmem$) based on peak memory bandwidth of CPU and GPU as in Equation \ref{eq:mem-convert}. CPU Xeon E5-2650L and GPU Nvidia Quadro K620 has a peak bandwidth 68 GB/s and 28.8 GB/s respectively. Both CPU and GPU platforms have two memory controllers. The bandwidth of one memory controller of GPU ($GB_{GPUcore}$) is equivalent to $\frac{28.8}{68}$ CPU counterpart, which is approximately half of the bandwidth of a CPU memory controller.
\begin{equation} \label{eq:mem-convert}
	nGPUmem = \frac{GB_{GPUcore}}{GB_{CPUcore}} \times nCPUmem
\end{equation}

The frequencies in REOH approach are represented by integer numbers as indexes. The increasing order of frequency indexes reflects the increasing oder of frequency values. For example, the experimented CPU has 8 frequencies (i.e., 1.2, 1.3, 1.4, 1.5, 1.6, 1.7, 1.8, 1.81GHz) represented by the numbers (i.e., 0, 1, 2, 3, 4, 5, 7, 8, respectively). The experimented GPU has one frequency (i.e., 1.73 GHz) represented by the number 6.
\subsection{Total energy consumption of heterogeneous systems}
In the REOH holistic approach, we target to optimize the total energy consumption of heterogeneous systems, including both static (idle) and dynamic energy of every platform in the system while the existing (homogeneous) approaches only consider the energy consumption of individual platform in isolation.
 
\begin{figure}[!t] \centering
	\resizebox{1\columnwidth}{!}{ \includegraphics{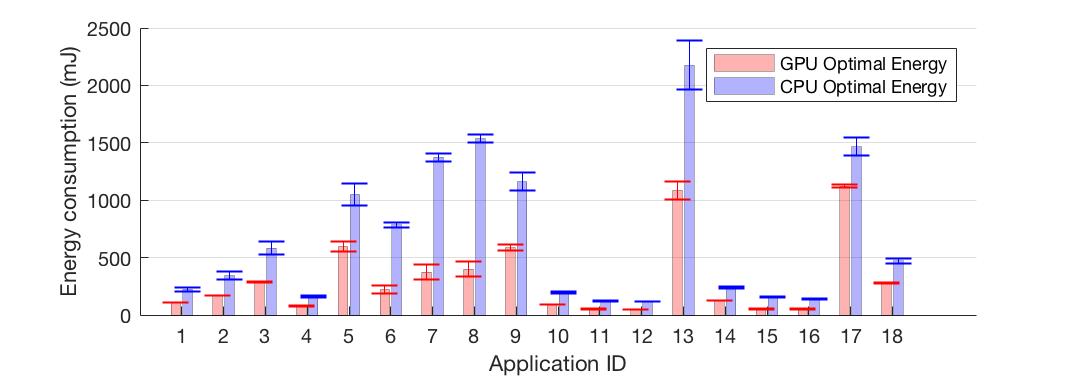}}
	\caption{Optimized energy consumption of CPU and GPU from homogeneous approach}
	\label{fig:CPU_GPU_comparison}
\end{figure}

\begin{figure}[!t] \centering
	\resizebox{1\columnwidth}{!}{ \includegraphics{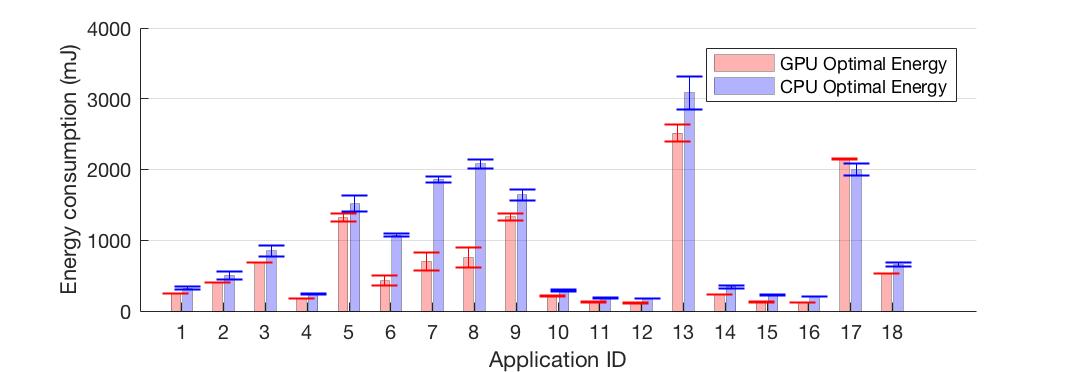}}
	\caption{Optimized energy consumption of CPU and GPU from the heterogeneous approach, which considers both static and dynamic energy of each platform}
	\label{fig:CPU_GPU_comparison_All_energy}
\end{figure}

%As demonstrated in Figures \ref{fig:CPU_GPU_comparison} and \ref{fig:CPU_GPU_comparison_All_energy}, the homogeneous approach %(i.e., LEO \cite{Mishra:2015}) 
%chooses to run application 17 on GPU to minimize energy consumption (cf. Figure \ref{fig:CPU_GPU_comparison}) since the energy consumption of the application on GPU is less than the energy consumption on CPU. However, the holistic approach %(i.e. REOH) 
%chooses to run application 17 on CPU (cf. Figure \ref{fig:CPU_GPU_comparison_All_energy}) when considering the energy consumption of the whole system consisting both CPU and GPU. 
Unlike the homogeneous approach that considers CPU energy and GPU energy in isolation, the holistic approach considers CPU energy and GPU energy together. It is because although the application runs on GPU (resp. CPU), idle CPU (resp. GPU) consumes energy as well (i.e., static energy). This is one of the reasons that makes the most energy efficient configurations from homogeneous approaches not always the most energy efficient configurations in heterogeneous systems. Figure \ref{fig:CPU_GPU_comparison} shows the optimal dynamic energy of CPU and GPU while \ref{fig:CPU_GPU_comparison_All_energy} shows the optimal total energy including static energy of the idle platform and dynamic energy of the running platform. The optimal configurations for each application from the two sets of data (i.e., the dynamic energy data and the total energy data) are not always the same. For example, from the dynamic energy data, running application 17 on GPU consumes less energy than running it on CPU while from the total energy data, running application 17 on CPU is more energy-efficient than on GPU.

%(Show two diagrams, one without the static and on with the static power and how the optimized setting can be different.)
%How the total energy is calculated affects the decision for selecting the platform and its configuration in the heterogeneous systems. 

The research question that the REOH approach wants to address is: which platform (CPU or GPU), together with its configuration, in a heterogeneous system is the most energy efficient for executing a given application. In our research context, when an application is executed by ones of the platforms (e.g., active platforms), the other platforms are in idle mode. The energy consumption of the active platforms includes their static and dynamic energy while the energy consumption of the idle platforms includes only their static energy. The total energy consumption of a whole heterogeneous system includes not only the energy of active platforms but also the energy of idle platforms as Equation \ref{eq:total-energy}. The energy consumption of active platforms includes static and dynamic energy while the energy consumption of idle platforms is the static energy. In Equation \ref{eq:total-energy}, the heterogeneous system has m platforms. The active platforms are platforms (1, 2, .., n) and the idle platforms are platforms (n+1, n+2, .., m). 
\begin{equation} \label{eq:total-energy}
%E^{total}=E^{static}_{platform1}+E^{dynamic}_{platform1}+E^{static}_{platform}
E^{total} = \sum\limits_{i=1}^n (E^{static}_{i}+E^{dynamic}_{i}) + \sum\limits_{j=n+1}^m E^{static}_{j}
%\sum\limits_{i=1}^n i^2 = \frac{n(n+1)(2n+1)}{6}
\end{equation}

In our heterogeneous system used for validating the REOH approach, there are two platforms CPU and GPU. If an application is run on CPU while GPU is idle, the total energy is computed as $E^{total}_{CPU}=E^{static}_{CPU}+E^{dynamic}_{CPU}+E^{static}_{GPU}$.  If an application is run on GPU while CPU is idle, the total energy is computed as $E^{total}_{GPU}=E^{static}_{GPU}+E^{dynamic}_{GPU}+E^{static}_{CPU}$. This is one of the improvements of REOH holistic approach compared to the existing (homogeneous) approaches.
\subsection{Application categories}
%Describe how to classify application, based on dwarf and algorithm characteristics, use table for explaining
We propose a selected set of applications for experimenting and devising a general training data set which can cover a wide range of communication and computation patterns.
%Based on [b], the impact factors are total execution time, class of application, problem size, programming language, I/O time, memory access time, CPU access time Study \cite{Asa06} 
\begin{table}
	%\resizebox{0.5\textwidth}{!}{\begin{minipage}{\textwidth}
	\caption{Application categories based on dwarf list}
	\label{table:Dwarf-list}
	\centering
	%\resizebox{\columnwidth}{!}{%
	\begin{tabular}{llllllll}
		\hline\noalign{\smallskip}
		\textbf{Dwarf} 	 &\textbf{Performance Limit\cite{Asa06} } &\textbf{Benchmark \cite{Rodinia}} 	\\						
			\noalign{\smallskip}\hline\noalign{\smallskip}					
Graph Traversal	&Memory Latency &B+Tree		 	   \\
&	&BFS 		 	   \\
			\noalign{\smallskip}\hline\noalign{\smallskip}						
				
Structured Grid	&Memory Bandwidth &HeartWall 		 	   \\
	& & Particle Filter  	   \\
			\noalign{\smallskip}\hline\noalign{\smallskip}						
Unstructured Grid	&Memory Latency	 &CFD Solver 	   \\
&	&Back Propagation 		 	   \\
		\noalign{\smallskip}\hline\noalign{\smallskip}						
		 Dense linear algebra	&Computation &LUD		 	   \\
		 		 	&	&kmeans	 	   \\
			\noalign{\smallskip}\hline\noalign{\smallskip}		
					 Sparse Matrix	&50\%Computation  &		 	   \\
				&50\% Memory Bandwidth &	 	   \\
					&(cf. Figure 9 in \cite{Asa06}) &	 	   \\
			\noalign{\smallskip}\hline\noalign{\smallskip}						
Dynamic Programming	&Memory Latency  &Path Finder 		 	   \\
&	&Needleman-Wunsch 		 	   \\
			\noalign{\smallskip}\hline\noalign{\smallskip}						
N-body		&Computation 	 	&LAVAMD    \\
		\noalign{\smallskip}\hline\noalign{\smallskip}
Spectral		&Memory Latency	&GPUDWT  	   \\
		\noalign{\smallskip}\hline\noalign{\smallskip}		
	\end{tabular}
\end{table}

%Describe the 5 chosen benchmarks and explain why choosing them
A training data set obtained offline is required by the probabilistic network approach. The main objectives of the training data set is to represent the wide range of computation and communication patterns and characteristics. In order to identify such varied set of patterns, we consider the pattern categories based on Berkeley dwarfs \cite{Asa06} and its corresponding benchmarks in the Rodinia benchmark suite \cite{Rodinia}. 

We summarize the dwarf list and their corresponded benchmarks based on their categories and characteristics as in Table \ref{table:Dwarf-list}. Each of the dwarfs has performance limit due to computation, memory bandwidth or memory latency as shown in the second column (e.g., Performance Limit). The third column shows the benchmarks belonging to the dwarf. 

There are several impact factors that affect the application performance and its optimization strategies such as algorithm design, execution configuration, control flow, memory types, memory access pattern and instruction count \cite{Feng:2012:ODW:2188286.2188341}. These factors are represented by three categories of performance limits: computation, memory bandwidth and memory latency \cite{Asa06}. In order to select the benchmarks that represent a wide range of applications behaviors, we choose a set of benchmarks that cover all three categories of the performance limits such as %LUD, 
Kmeans, BFS, Particle Filter and CFD. The four benchmarks %LUD, BFS, Particle Filter and CFD 
belong to the first four dwarfs in Table \ref{table:Dwarf-list}. %Kmeans belongs to Dense Linear Algebra but it is the benchmark has performance limit at the borderline of memory bandwidth and computation, depending on which platform it is run on. E.g., Kmeans has memory bandwidth as performance limit on CPU and computation as performance limit on GPU \cite{Feng:2012:ODW:2188286.2188341}.

%Describe why choosing Rodinia benchmarks, why opencl
We chose Rodinia \cite{Rodinia} benchmarks to validate our approach because it provides implementations for a variety of platforms (e.g., CPU and GPU) and programming models (e.g., OpenCL, CUDA, OpenMP). Among the supported programming models of Rodinia, OpenCL implementations are selected since OpenCL library is supported on a various architectures such as CPU, GPU and accelerators. 
%Therefore, for heterogeneous systems including different architectures, Rodinia benchmarks is chosen for conducting experiments to devise the training data set and evaluate the proposed energy optimization approach. 

Moreover, the problem size can also impact the benchmark performance and its optimization strategy \cite{Feng:2012:ODW:2188286.2188341, 7823853}. For each chosen benchmarks, we also select a set of input that covers a varying range of benchmark patterns. %%%%%%%%%%

%The selected input is further explained in Section \ref{training_data}. 
The selected input was generated using the data generators from Rodinia, in which the sample sizes were chosen to grow exponentially to cover various range of input sizes. BFS has input graphs with sizes varying from 512kB to 8MB. CFD experiments are conducted with only three input sizes due to the unavailability of input generator and limited input provided by Rodinia.
%removed for CR version 
%Its three input sizes, however, still satisfy our requirements: the size ranges ensures that the benchmark running time is at least 2 ms for each input and each configuration and the total experiment time is reasonable long not to add the heat overhead to the energy measurement. 
%LUD has input sizes as the dimension of matrixes varying from 512 to 8192 elements. 
Kmeans has the input generating from two parameters: the number of objects and the number of features. For instance, in Table \ref{table:applicationID}, the input name 1000\_34 means there are 1000 objects and each object has 34 features \cite{NIPS2009_3812}. Particle Filter has the input generating from three parameters as its three dimensions. For instance, the input name 128\_10\_1000\_dp means that the input dimensions is 128x128x10 with 1000 particles and particles are double type \cite{cclw:gpu_cfd_unstructured:2009}. For each input size and configuration, each benchmark is performed five times and the measurement of average and deviation values are stored in training data set.
%(e.g., 5 days for each benchmark). 
%Samples from benchmarks that ran over 10 seconds were necessary in order to achieve accurate measures using the Leo framework. The input sizes were chosen to  t a time frame. 1 second to two magnitudes higher were deemed appropriate due to the number of permutations.
%The selected input was generated using the data generators from Rodinia, sample sizes were chosen to grow logarithmically with the exception where increasing an order of magnitude would require too much time.

\begin{table}
	%\resizebox{0.5\textwidth}{!}{\begin{minipage}{\textwidth}
	\caption{Application details}
	\label{table:applicationID}
	\centering
	%\resizebox{\columnwidth}{!}{%
	\begin{tabular}{llllllll}
		\hline\noalign{\smallskip}
		\textbf{Application ID} 	 &\textbf{Benchmark} &\textbf{Input} 	\\						
		\noalign{\smallskip}\hline\noalign{\smallskip}						
		1	&BFS &graph1M		 	   \\
		2	&BFS &graph2M		 	   \\
		3	&BFS &graph4M		 	   \\
		4	&BFS &graph512k		 	   \\
		5	&BFS &graph8M		 	   \\
		\noalign{\smallskip}\hline\noalign{\smallskip}		
		6	&CFD  &fvcorr.domn.097K		 	   \\
		7	&CFD  &fvcorr.domn.193K		 	   \\
		8	&CFD  &missile.domn.0.2M	 	   \\
		\noalign{\smallskip}\hline\noalign{\smallskip}					
		9	&Kmeans &1000000\_34		 	   \\
		10	&Kmeans &100000\_34		 	   \\
		11	&Kmeans &10000\_34		 	   \\
		12	&Kmeans &1000\_34		 	   \\
		13	&Kmeans &3000000\_34		 	   \\
		\noalign{\smallskip}\hline\noalign{\smallskip}								
		14	&ParticleFilter &128\_10\_100000\_dp 		 	   \\
		15	&ParticleFilter &128\_10\_10000\_dp 		 	   \\
		16	&ParticleFilter &128\_10\_1000\_dp 		 	   \\
		17	&ParticleFilter &128\_2500\_10000\_dp 		 	   \\
		18	&ParticleFilter &128\_500\_10000\_dp 		 	   \\
		%19	&LUD &1024		 	   \\
		%20	&LUD &2048 		 	   \\
		%21	&LUD &4096 		 	   \\
		%22	&LUD &512 		 	   \\
		%23	&LUD &8192 		 	   \\
		\noalign{\smallskip}\hline\noalign{\smallskip}		
	\end{tabular}
\end{table}  

\section{Energy Saving - Experimental Results}
\label{energy_saving}
%\subsection{Validation of the platform-adaptive tuning approach}
In this section, we validate the REOH approach by experimental study: how close to the optimal configuration (by the brute-force approach) the configuration by the REOH approach is. The optimal configuration means the best platform and its best setting in term of energy consumption. The REOH approach predicts the best configurations (i.e., the best platform and its best setting in term of energy consumption) based on the training data and sampling data. 

\subsection{Devise training data and sampling data}
The training data was devised by conducting the experiments to measures energy consumption of 18 applications (each application is a combination of a benchmark and an input) on all available configurations of two platforms (i.e., 384 configurations of CPU and 9 configurations of GPU) in the targeted heterogeneous system (cf. Table \ref{table:applicationID}). The 384 configurations of CPU are the combination of 24 cores, 8 frequencies and 2 memory controllers. The CPU configurations (i.e., the combinations of cores, frequencies, memory controllers) are set by using \textit{cpufrequtils} package and \textit{numactl} library. The 9 configurations of GPU are the workgroup sizes assigned to applications, such as 1, 2, 4, 8, 16, 32, 64, 128, 256 work units. %The workgroups have two dimensions, which are equal to square sizes such as $1\times1, 2\times2, 4\times4, 8\times8, 16\times16$. 
%The workgroup sizes should be less than the number of cores of the GPU which is 384 cores. %add after OPTIM comments
Each application was run five times for each configuration and the mean and standard deviation values of measured performance and consumed energy are stored. Note that the minimum number of cores (respectively memory controller) is one in order to ensure that the application always completes in a finite amount of time. 

The sampling data is obtained by running a given application with sample configurations and measuring its performance time and energy consumption. In our validation, sample configurations are chosen randomly. 
%, except the minimum configuration (i.e., the minimum number of cores, the lowest frequency, the minimum number of memory controllers) and maximum configurations (i.e., the maximum number of cores, the highest frequency, the maximum number of memory controllers) on both platforms. 
%add after OPTIM comments
%from CPU and 2 samples (i.e., 1 and 256 work units) from GPU. The presence of GPU samples in the training set increases the prediction accuracy.

%Each  were performed 5 times for 5 different input sizes. Samples from benchmarks that ran over 10 seconds were necessary in order to achieve accurate measures using the Leo framework. The input sizes were chosen to fit a time frame. 1 second to two magnitudes higher were deemed appropriate due to the number of permutations.The selected input was generated using the data generators from Rodinia, sample sizes were chosen to grow logarithmically with the exception where increasing an order of magnitude would require too much time.
\subsection{Approach validation}
\label{sec:Approach validation}
Based on the training data and sampling data, the probabilistic model is applied to estimate the energy consumption of the remaining configurations (namely, all possible configurations except for sample configurations). Noted that when sampling an application $A$, $A$'s data is removed from the training data set. From the estimated energy consumption of all configurations, the best configuration which consumes the least energy is selected. 

We compares the result of the REOH approach with the LEO approach \cite{Mishra:2015}, the state-of-the-art (homogeneous) approach based on a similar probabilistic model. REOH approach is applied on a heterogeneous system with both CPU and GPU data while LEO approach is applied on homogeneous system (i.e., either on CPU platform with CPU data or GPU platform with GPU data). The details (i.e., data from which platform and data size) of training and sampling set for each approach are summarized in Table \ref{table:trainingsampling}. 

The probabilistic approach uses regression diagnostics (i.e., regstats function) \cite{regstats} with full quadratic \cite{x2fx} as an input model. For REOH and LEO-CPU prediction, the regstats function has 3 predictors (i.e., the number of cores, the frequency index and the number of memory controllers) which creates 10 (i.e., $\frac{(3+1)\times(3+2)}{2}$) predictor variables \cite{x2fx}. The model for REOH and LEO-CPU, therefore, requires at least 10 observations (i.e., the number of sampling data). Since the considered GPU has less than 10 configurations, we only use one predictor (i.e., workgroup size) for the regression function when applying the probabilistic approach for GPU platform with GPU data only. The model for LEO-GPU requires at least 3 sampling data.

The prediction was performed with the total number of samples varying from 10 (the minimum samples requirement) to 50 samples. The accuracy of the model increases when the number of sample increases to 15. After reaching 15 samples, the accuracy of the model does not significantly changed when taking more samples. Therefore, we choose to sampling 15 data on 15 configurations when performing model prediction with REOH and LEO-CPU approach. For LEO-GPU, we choose the number of sampling data as 3. 

In this validation, we compare the most energy-efficient configuration by the REOH approach for a heterogeneous system containing a CPU and a GPU to the most energy-efficient configurations by the LEO approach for a homogeneous system with a CPU platform and the most energy-efficient configurations by the LEO approach for a homogeneous system with a GPU platform. Moreover, we also compare the REOH results with the optimal results by the brute-force approach that has all measured data of all platforms (i.e., CPU and GPU) available. The brute-force approach always choose the optimal configuration.
\begin{table}
	%\resizebox{0.5\textwidth}{!}{\begin{minipage}{\textwidth}
	\caption{Training and sampling data for each approach}
	\label{table:trainingsampling}
	\centering
	%\resizebox{\columnwidth}{!}{%
	\begin{tabular}{llllllll}
		\hline\noalign{\smallskip}
		\textbf{} 	 &\textbf{Training Data} & &\textbf{Sampling Data} &	\\	
		\noalign{\smallskip}\hline\noalign{\smallskip}
		\textbf{Approach} 	 &\textbf{Platform} &\textbf{Size}  &\textbf{Platform} &\textbf{Size}\\					
		\noalign{\smallskip}\hline\noalign{\smallskip}						
		LEO-CPU &CPU &384x18 &CPU &15x1		 	   \\
		\noalign{\smallskip}\hline\noalign{\smallskip}		
		LEO-GPU &GPU &9x18 &GPU &3x1		 	   \\
		\noalign{\smallskip}\hline\noalign{\smallskip}					
		REOH &CPU+GPU &393x18 &CPU &15x1		 	   \\
		%Brute Force &CPU+GPU &393x18 &CPU &393x1		 	   \\
		\noalign{\smallskip}\hline\noalign{\smallskip}		
	\end{tabular}
\end{table}
%It is noted that the probabilistic approach uses regression diagnostics (i.e., regstats function) \cite{regstats} with full quadratic \cite{x2fx} as an input model. The regstats function has 3 predictors (i.e., the number of cores, the frequency index and the number of memory controllers) which creates 10 (i.e., $\frac{(3+1)\times(3+2)}{2}$) predictor variables \cite{x2fx}. The model, therefore, requires at least 10 observations (i.e., the number of sample configurations). Since the considered GPU has less than 10 configurations, we only use one predictor (i.e., workgroup size) for the regression function when applying the probabilistic approach for GPU platform with GPU data only. 

\begin{figure}[!t] \centering
	\resizebox{1\columnwidth}{!}{ \includegraphics{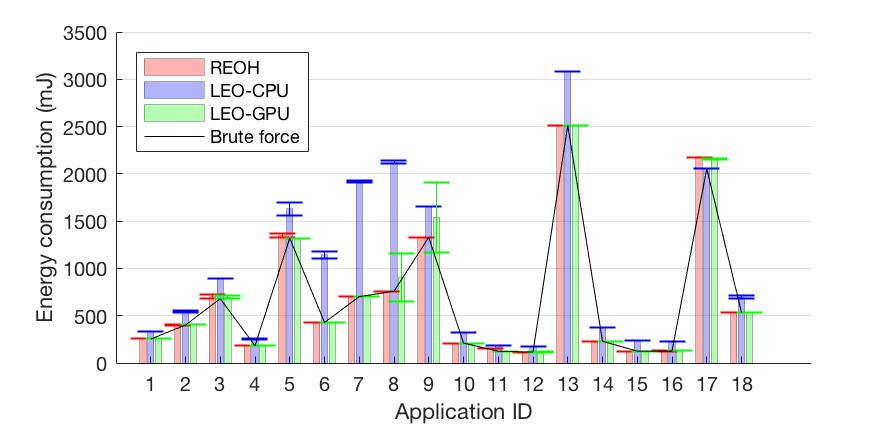}}
	\caption{Energy comparison of the four approaches: REOH, LEO-CPU, LEO-GPU and Brute Force}
	\label{fig:Energy_compare_1}
\end{figure}

\begin{figure}[!t] \centering
	\resizebox{1\columnwidth}{!}{ \includegraphics{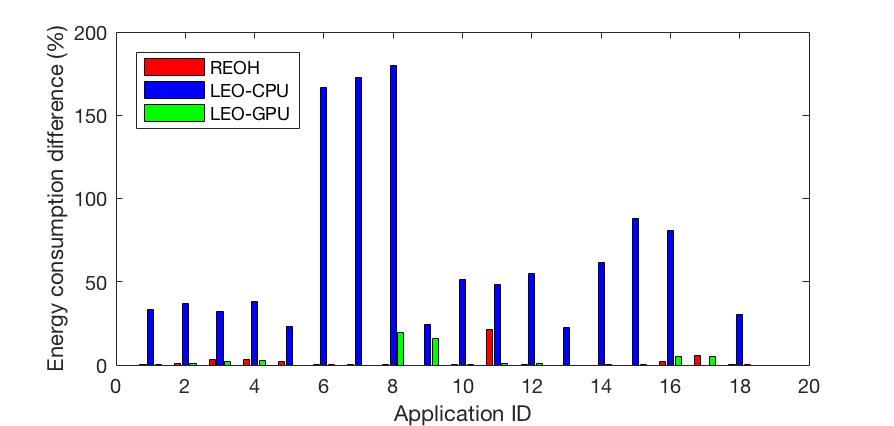}}
	\caption{Percentage of the differences on energy consumption of REOH, LEO-CPU and LEO-GPU approach compared to Brute Force approach}
	\label{fig:Energy_percentage_1}
\end{figure}

Figure \ref{fig:Energy_compare_1} shows the energy consumption (in mJ) of the configurations selected by the four approaches for 18 applications and Figure \ref{fig:Energy_percentage_1} shows the energy consumption difference between the three approaches (LEO-CPU, LEO-GPU \cite{Mishra:2015} and REOH) and the Brute-force approach. The list of applications and their ID are summarized in Table \ref{table:applicationID}.

The results shows that for 17 out of 18 applications, the the REOH approach predicts the close results to LEO-GPU approach and the Brute Force approach (up to 0.9\% more energy consumption to LEO-GPU and within 5.7\% deviation to Brute Force) except application 11. 
%Unlike other applications where the number of cores significantly affect the energy consumption, application 11 has energy consumption from CPU experiments where the most energy-efficient configurations only depend on the frequency and not the number of cores. 
Unlike other applications where the performance increases when the number of cores increases, application 11 has the performance increased in the first 12 cores and decreased in the second 12 cores as shown in its experimental data (note that the platform has two 12-core CPUs). Application 11 has a different performance pattern than other applications which leads to the less precise prediction of REOH on application 11. 
%This is because application 11 has small problem size and the system overhead has more impact. Therefore, for application 11, it can not be predicted that when the number of cores is big (e.g., GPU configurations), the energy consumption can be reduced. 
%Application 11 has different performance pattern than other applications and that leads to the less precise prediction of REOH. 
%To compare with Brute Force approach, the results by the REOH approach is less than 5\% difference from the Brute-force except application 12. This happens with the LEO-CPU approach as well.
%The reason is that the most energy efficient configurations of application 12 falls into configurations with less core and the highest frequencies of CPU. For application 12, the configurations with highest frequency are more energy-efficient and the number of cores does not significantly affect the energy consumption. It does not predict that when the number of cores is big (e.g., GPU configurations), the energy consumption can be reduced.
 
REOH also predicts better results than LEO-CPU except application 17.
%Application 17 has the most energy-efficient configuration on CPU and therefore, LEO-CPU approach considering only CPU data as a training set provides more accurate prediction than REOH. However, REOH still achieves the same result as LEO-GPU for application 17.
LEO-CPU approach has better prediction only on the application 17: 5.7\% less energy consumption than the REOH approach. Application 17 has the best configuration on the CPU platform and the LEO-CPU approach, which considers only CPU data, is expected to be more accurate. However, its energy difference on the CPU platform between LEO-CPU and REOH approaches is marginal. 
Even though REOH approach predicts a configuration with higher energy consumption than LEO-CPU approach at application 17, its energy consumption is also within 5.7\% of the optimal energy consumption by the brute-force approach (cf. Figure \ref{fig:Energy_percentage_1}).  

%Application 6 has the higher difference percentage because the application itself has high deviation of measured energy values in its experiments as shown in Figure \ref{fig:CPU_GPU_comparison_All_energy}, which results in higher deviation when selecting the most energy-efficient configuration. 
%This happens with the LEO approach as well: for application 6, the energy consumption by LEO approach is 170\% higher than the energy consumption by the brute-force approach. 
%For the 21 other applications (except application 17), the energy consumption by the LEO approach is from 20\% up to 195\% higher than the energy consumption by the brute-force approach, which is much higher than the energy consumption by REOH approach (cf. Figure 4).
The results have confirmed that the REOH approach can use the training set from selected applications to predict competitive configurations (within 5.7\% of the optimal in 17 applications) in term of energy consumption. %The results also confirms that the REOH approach provides better accuracy compared to the LEO approach, the state-of-the-art approach based on a probabilistic model.
Moreover, the REOH approach only needs 15 samples from CPU data to predict the most energy-efficient configuration while LEO requires two predictions on data from two separate platforms, either CPU or GPU data. The total number of samples when using LEO approach is $15+3=18$, which is 20\% more sampling numbers as compared to REOH approach. By using REOH approach, the system is beneficial in two ways: not sampling GPU data and save 17\% (i.e., $\frac{3}{15+3}$) the number of sampling runs.  

\section{Energy-optimizing Runtime Framework}
\label{runtime_app}
%Draw a diagram of applicationn design with 4 components: devise training set, get samples, prediction and identify the best configuration, run the executable file. Explain each of the components from the design view
%\section{System Description}
%A runtime prototype has been implemented to provide users with optimal configurations for a given application and n runtime tool to run their given executable file on the most energy-efficient of the targeted heterogeneous system. The code is available at: .
Based on the new REOH approach, an open-source runtime framework has been developed to provide users with an energy-efficient system configuration for a given executable running on a heterogeneous system. The framework is publicly available at: https://github.com/uit-agc/REOH
\subsection{Framework design}
\begin{figure}[htp]
	\centering
	\includegraphics[width=0.5\textwidth]{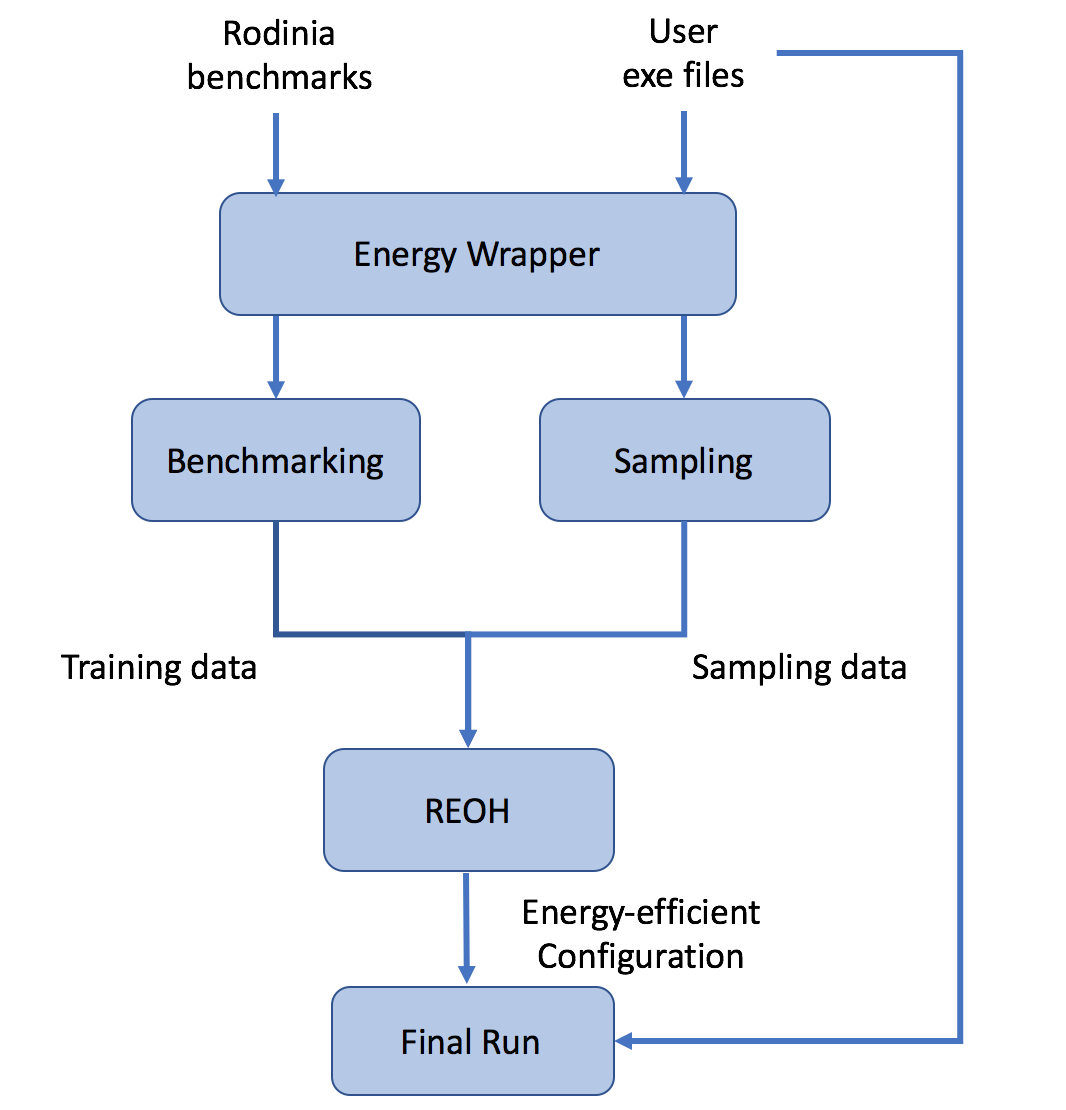}
	\caption{Prototype Overview}
	\label{Fig:prototype}
\end{figure}

%\subsubsection{Prototype design}
Figure \ref{Fig:prototype} shows an overview of our framework.
%\subsection{Rodinia}
\paragraph{Energy Wrapper}
The energy wrapper consists of an executable that is responsible for setting platform configurations and measuring energy and execution time of a given application. Each application should provide two executables: one for the CPU platform and one for the GPU platform, assuming the the underlying heterogeneous system consists of CPU and GPU platforms. Time and energy measurement were performed using MeterPU \cite{REPARA15}, instantiated with Intel PCM for CPU and Nvidia NVML for GPU. The executables are executed using the POSIX \texttt{system()} command. % thus the overhead of performing a system call to execute a new child process as well as setup is included in the measurements.
\paragraph{Benchmarking}
The module is to obtain the training data for a given heterogeneous system by executing the energy wrapper module over all 18 applications (cf. Table \ref{table:applicationID}) for all system configurations. This step only needs to perform once for different workloads. %A separate testing component were implemented to produce the formated output for specific applications using a fixed set of samples.
\paragraph{Sampling}
The sampling is performed by executing the energy-wrapper for user executables on sample configurations. This module is to provide the sampling data in order to estimate the energy consumption of the executables on all configurations. This step is performed for every given application and its executables from users. %A separate testing component were implemented to produce the formated output for specific applications using a fixed set of samples.

The output data of both the Benchmarking and Sampling module is converted to the appropriate format using the scripts provided in this framework. During transformation, we also add static energy consumed by CPU and GPU.
%\paragraph{Static Energy Measurement}
The static energy were measured by recording the energy measurements over 20 seconds for each platform using MeterPU\cite{REPARA15}. This was done once to measure the the static power of each platform in the heterogeneous system. The static power are stored for later use.
\paragraph{REOH}
The energy-optimizing module estimates the energy consumption of all configurations of the heterogeneous system based on the training data set and sampling data set. Then it provides a appropriate energy-efficient configuration to run the given application.
\paragraph{Final Run}
From the configuration provided by REOH module, the Final Run module runs the appropriate executable file (e.g., executable file for CPU or GPU) on the provided configuration and measure its energy consumption.
%\section{Experimentation}
%\section{Implementation}
%\section{Input}
%Each benchmark were performed 5 times for 5 different input sizes. Samples from benchmarks that ran over 10 seconds were necessary in order to achieve accurate measures using the Leo framework. The input sizes were chosen to fit a time frame. 1 second to two magnitudes higher were deemed appropriate due to the number of permutations.
%The selected input was generated using the data generators from Rodinia, sample sizes were chosen to grow logarithmically with the exception where increasing an order of magnitude would require too much time.
%\subsection{Framework Validation }
\subsection{Implementation details}
In order to measure energy consumption of Rodinia benchmarks, two main modifications have been made.
i) Changing the OpenCL setup to target our specific platforms. ii) Reading work-group sizes from environment for GPU. Executables were produced for each platform, which were then sent to energy-wrapper module.

All benchmarks were performed on a 24-core Intel(R) Xeon(R) CPU E5-2650L v3, with 64 GB RAM, running Linux 3.10.0. For GPU experiments an Nvidia Quadro K620 was used. The prototype was implemented using C, C++ and OpenCL for the Rodinia benchmarks and Energy Wrapper component. Data transformation and benchmarking tools were implemented using Bash and AWK.
%\subsubsection{Validation results}
%\begin{figure}[!t] \centering
%	\resizebox{0.8\columnwidth}{!}{ \includegraphics{diagrams/Framework_validation.jpg}}
%	\caption{Energy consumption of the proposed approach compared to }
%	\label{fig:framework-validation}
%\end{figure}   

\section{Related Work}
\label{related_work}
The related work to this study is summarized in Table \ref{table:auto-tuning-framework}. The summary shows that the previous approaches are either for tuning the code variants (i.e., implementations of an algorithm or an application) and therefore restrict to specific applications \cite{OSKI2005, 6877283};
or for tuning configurations for a general application but only applicable to homogeneous system with a single unit type (i.e., CPU or GPU) \cite{POET, Powercap, Mishra:2015, 6877283, 6413638, 6877247, Nath:2015:CPM:2830772.2830826, 8327055, Wang:2017:GPE:3152042.3152066}.  %Study \cite{6877283} tune the code variants according to GPU devices. The approach is however, also restrict to a specific set of applications and is not for a general applications.
%OpenCL \cite{OPENCL} has become the de-facto data parallel programming model for parallel devices in today's high-performance supercomputers. OpenCL was designed with the goal of guaranteeing program portability across hardware from different vendors. However, achieving good performance is hard, requiring manual tuning of the program and expert knowledge of each target device.
%In this paper we consider a data parallel compiler transformation --- thread-coarsening --- and evaluate its effects across a range of devices by developing a source-to-source OpenCL compiler based on LLVM. We thoroughly evaluate this transformation on 17 benchmarks and five platforms with different coarsening parameters giving over 43,000 different experiments. We achieve speedups over 9x on individual applications and average speedups ranging from 1.15x on the Nvidia Kepler GPU to 1.50x on the AMD Cypress GPU. Finally, we use statistical regression to analyse and explain program performance in terms of hardware-based performance counters.
%The group of studies \cite{ATLAS2000, LAPACK, SPIRAL, OSKI2005} provides auto-tuning framework for stencils kernels to utilize architectural resources and allow performance portability across diverse architecture. However, they mainly focus on tuning specific implementations or algorithms and are not applicable to a general application.

Study about PowerCap \cite{Powercap} chooses the most suitable settings for energy efficiency but still meet the performance requirement. It operates based on feedback, then observes, decide and act. The approach requires feedback on a certain platform (e.g., CPU) which is not applicable for heterogeneous systems. POET \cite{POET} also chooses the system configuration to meet energy requirement based on feedback and controllers. However, POET is also applicable to homogeneous system with a single unit type (e.g., CPU). 

%Nitro framework \cite{6877283} and OSKI \cite{} tunes the code variants according to GPU devices. The approach is however, also restrict to a specific set of applications and is not for a general applications.
%Lee et al. \cite{Guevara:2014:MMM:2584468.2541258} focuses on how to match the user requests to the architecture profile and distribute tasks from a set of applications to a heterogeneous datacenter without affecting the quality of service of the applications. 
%proposes a solution to distribute tasks from a set of applications to datacenter containing heterogeneous architectures. 
There are a group of studies that provides power and performance models for GPU and GPGPU to predict the most energy-efficient DVFS configuration of GPU to run an application \cite{6877283, 6413638, 6877247, Nath:2015:CPM:2830772.2830826, 8327055, Wang:2017:GPE:3152042.3152066, 7920860}. The models, however, are for the considered GPUs and not for heterogeneous systems. 
Studies \cite{7360199, 7501903, 7863726} develop frameworks for workload partitioning on a type of heterogeneous systems (i.e., APUs) but they are mainly focus on improving performance instead of energy-efficiency.

There are two approaches for heterogeneous systems in the Table \ref{table:auto-tuning-framework}: GreenGPU \cite{6337630, MA201621} and the market mechanism \cite{Guevara:2014:MMM:2584468.2541258}. GreenGPU \cite{6337630, MA201621} targets iterative applications (i.e., applications have several iterations where the next iteration execution time can be predicted based on the current iteration) which is different from REOH (i.e., REOH is applicable for general applications and requires no prior knowledge of applications). %The algorithm of MPC \cite{7920860} does not use the off-line training data but instead, inlcudes a chain of kernel invocations as the learning steps. The MPC algorithm requires the understanding of application details while REOH does not need prior knowledge of applications. 
The market mechanism \cite{Guevara:2014:MMM:2584468.2541258} requires three analysis and optimization phases to match the user profile to the architecture profile and distribute the application to the hardware. %The process is efficient for a large set of applications but is complicated for a single application. 
%The first phase of estimating task arrival distribution uses Gaussian distribution.
%Their study distribute the application to the hardware to achieve the least cost of the whole set of applications
The chosen configuration from the market mechanism \cite{Guevara:2014:MMM:2584468.2541258} is at datacenter level (i.e., its targeted configuration is a mix of CPUs and microprocessors) while the chosen configuration of REOH is at platform level (i.e., REOH configuration is a combination of the number of cores, frequency and memory controllers).
%to balance the resource allocation among the whole set of applications and therefore, might not be the most energy-efficient configuration for given applications. The work presented in this paper is to find the most energy-efficient hardware configuration for a given application running on a heterogeneous system.

This study is inspired by LEO framework \cite{Mishra:2015}. LEO chooses the best system configuration depending on the application and its input. This approach uses probabilistic graphical models to estimate the energy consumption of applications. However, LEO only considers a set of configurations of a CPU-based homogeneous system. The present approach (this study) apply probabilistic network approach to identify the most energy-efficient configuration for an application running on heterogeneous systems and tune the configurations in runtime.  
%Describe/ summarize the auto-tuning approaches in literature.  

\section{Conclusion}
\label{conclusion}
This study has proposed and validated REOH, a new holistic approach using probabilistic model to predict and select the optimal configurations in term of energy consumption of heterogeneous systems for a given application. This study has demonstrated that REOH can achieve almost optimal energy consumption (within 5.7\% of the optimal energy consumption by the brute-force approach) while saving the energy consumption of 17\% less sample runs. Based on the REOH approach, a runtime framework for executing given executables energy-efficiently is developed and provided as open source software for scientific purposes.  

\bibliographystyle{ieeetr}
\bibliography{./REO}

\end{document}